\documentclass[aps,prb,supersciptaddress,twocolumn,showpacs]{revtex4}%

\usepackage{amsfonts}
\usepackage{amsmath}
\usepackage{amssymb}
\usepackage{graphicx}%

\begin{document}
\title{Vortex phase diagram of Ba(Fe$_{0.93}$Co$_{0.07}$)$_{2}$As$_{2}$ single crystals}
\author{R.~Prozorov}
\email[Corresponding author: ]{prozorov@ameslab.gov}
\affiliation{Ames Laboratory and Department of Physics \& Astronomy, Iowa State University, Ames, IA 50011}
\author{N.~Ni}
\affiliation{Ames Laboratory and Department of Physics \& Astronomy, Iowa State University, Ames, IA 50011}
\author{M.~A.~Tanatar}
\affiliation{Ames Laboratory and Department of Physics \& Astronomy, Iowa State University, Ames, IA 50011}
\author{V.~G.~Kogan}
\affiliation{Ames Laboratory and Department of Physics \& Astronomy, Iowa State University, Ames, IA 50011}
\author{R.~T.~Gordon}
\affiliation{Ames Laboratory and Department of Physics \& Astronomy, Iowa State University, Ames, IA 50011}
\author{C.~Martin}
\affiliation{Ames Laboratory and Department of Physics \& Astronomy, Iowa State University, Ames, IA 50011}
\author{E.~C.~Blomberg}
\affiliation{Ames Laboratory and Department of Physics \& Astronomy, Iowa State University, Ames, IA 50011}
\author{P.~Prommapan}
\affiliation{Ames Laboratory and Department of Physics \& Astronomy, Iowa State University, Ames, IA 50011}
\author{J.~Q.~Yan}
\affiliation{Ames Laboratory and Department of Physics \& Astronomy, Iowa State University, Ames, IA 50011}
\author{S.~L.~Bud'ko}
\affiliation{Ames Laboratory and Department of Physics \& Astronomy, Iowa State University, Ames, IA 50011}
\author{P.~C.~Canfield}
\affiliation{Ames Laboratory and Department of Physics \& Astronomy, Iowa State University, Ames, IA 50011}

\date{8 October 2008}

\begin{abstract}
Detailed measurements of the global and local electromagnetic properties of
Ba(Fe$_{0.93}$Co$_{0.07}$)$_{2}$As$_{2}$ single crystals are reported. Analysis of
the irreversible magnetic response provides strong evidence for similar vortex
physics in this Fe-based pnictide superconductor to the 
high-$T_{c}$ cuprates, such as Y-Ba-Cu-O or Nd-Ce-Cu-O. In particular, we have found a
nonmonotonic "fishtail" magnetization in $M\left(H,T=const\right)$ loops and
its signature is also present in $M\left(H=const,T\right)$ scans. The supercurrent density is evaluated by using several techniques, including direct transport measurements. At 5 K we estimate its value to be a moderate $j \approx 2.6 \pm 0.2 \times 10^5$ A/cm$^2$. Analysis of the magnetic relaxation is consistent with the collective pinning and creep models (weak pinning and fast creep) and suggests a crossover from the collective to the plastic creep regime in fields exceeding the value corresponding to the maximum in fishtail magnetization.

\end{abstract}


\maketitle

\section{Introduction}

The behavior of type-II superconductors in applied magnetic fields has been a
subject of ongoing interest due to obvious applications as well as
deep connections to the microscopic properties of superconductors and
mechanisms of superconductivity (for some reviews, see Refs.~[\onlinecite{Campbell1972,Dew-Huges1974,Tinkham1975,Ullmaier1975,blatter94,Brandt1995,Yeshurun1996}]). The situation is complicated in type-II superconductors because magnetic fields penetrate the sample in the form of Abrikosov vortices~\cite{Abrikosov57English,Abrikosov57Russian} whose
collective behavior determines the macroscopic electromagnetic response. The interaction between vortices and local modulations of the superconducting properties leads to pinning, which together with thermal fluctuations results in spatio-temporal variations of the vortices (magnetic induction) and  therefore the irreversible magnetic properties. Strong thermal fluctuations and weak pinning lead to complex $H-T$ diagrams with various crossovers in the static and dynamic electromagnetic response of vortex matter. 

In the case of layered superconductors, such as the high-$T_{c}$ cuprates and ET organics, a non-monotonic dependence in the field dependent magnetization is observed (see Refs.~[\onlinecite{Zuo1996,Pissas2000,Mikitik2001,Xu2008}]). This is called the "fishtail" or peak-effect.  The latter term is ambiguous since a different nonmonotonic behavior, usually closer to the second critical field, is observed in low-$T_{c}$ superconducting alloys~\cite{DeSorbo1964,Campbell1972}.   Examples of this can be found in materials such as NbSe$_{2}$~\cite{Banerjee2000} and neutron-irradiated V$_{3}$Si~\cite{Meier-Hirmer1985} and MgB$_{2}$.\cite{Zehetmayer2004} Another prominent feature of high-$T_{c}$
superconductors is a very large rate of magnetic relaxation (giant flux creep, reported by Yeshurun and Malozemoff~\cite{Yeshurun1988}) and its complicated time, magnetic field, and temperature dependence.~\cite{blatter94,Brandt1995,Yeshurun1996} While the mechanisms of the magnetic relaxation and fishtail feature in the cuprates are still being discussed, several self-consistent microscopic numerical simulations and phenomenological models have been proposed.  Some of these
 have been able to explain the observed effects in many classes of unconventional superconductors. Magnetic relaxation with an emphasis on high-$T_{c}$ superconductors is reviewed by Yeshurun, Malozemoff, and Shaulov.~\cite{Yeshurun1996} A detailed analysis of the pinning strength and the barriers for magnetic relaxation in the case of weak collective pinning and creep
is given by Blatter \textit{et al.}~\cite{blatter94} and also by Brandt.~\cite{Brandt1995}

In disordered granular and polycrystalline materials much of the response comes from extrinsic factors, such as granularity and significant disruption of supercurrent flow between the grains. To study intrinsic magnetic
properties that can be related to the mechanism of superconductivity, such as anisotropy and basic length scales (coherence length, $\xi$, and London penetration depth, $\lambda$), one must turn to single crystals. A vast amount of literature on this subject clearly shows that different classes of superconductors exhibit unique static and dynamic magnetic responses. This allows us to discuss the similarities and differences in vortex behavior and to look for the connections between different classes of materials.  Vortex phase diagrams that show various transitions, such as changes in vortex lattice symmetry, dimensionality or
mobility.  Pinning strength and magnetic dynamics are particularly useful for such comparisons and determining applications. Recent examples are CeCoIn$_{5}$~\cite{Bianchi2008} and MgB$_{2}$.~\cite{Welp2003}

So far, only a limited amount of information regarding the vortex behavior in single crystals of Fe-based pnictide superconductors is available. In polycrystalline samples, the magnetic behavior is significantly affected by extrinsic factors, such as grain morphology, surface roughness, and inter-grain voids and interfaces. In these materials, it has been found that some of the grains are not even superconducting~\cite{Fe_Prozorov2008}. Although the superconducting granularity can be estimated from magneto optical imaging,  and some important parameters can still be extracted~\cite{Yamamoto2008a,Jaroszynski2008b}, the polycrystalline structure does complicate the analysis. For example, a signature of nonmonotonic $M(H)$ has been reported in polycrystalline wires of SmFeAsO$_{0.8}$F$_{0.2}$~\cite{Chen2008e} and could be related to the fishtail feature of this work, but it could also be due to the physics of inhomogeneous multi-phase materials. Furthermore, severe effects of inter-grain weak-links did not allow for quantitative analysis of the intrinsic properties.  

The first measurements of magnetization and relaxation in individual single crystals of the oxypnictide superconductor NdFeAsO$_{1-x}$F$_{x}$ have revealed the vortex physics to be surprisingly reminiscent of cuprate superconductors, in particular the existence of a large magnetic relaxation rate and relatively weak pinning.~\cite{Fe_Prozorov2008} Later torque measurements on the similar Sm-based system have lead to the conclusion that the electromagnetic anisotropy is large and quite temperature dependent.~\cite{Weyeneth2008a} More recent high-field torque measurements on similar crystals have shown a much smaller anisotropy.~\cite{Balicas2008} The superconducting crystals based on the parent RFeAsO compound (1111 system, where R is a rare earth element), are small and difficult to obtain (via high-pressure synthesis) and study. On the other hand, large crystals based on the related oxygen-free parent compound AFe$_{2}$As$_{2}$ (122 system, where A is an alkaline element), can be grown in flux at
ambient pressure (see, e.g., A=Ba\cite{Ni2008}, Sr\cite{Yan2008} and Ca\cite{Ni2008b}). Both A and Fe sites can be doped to achieve superconductivity with holes or electrons as carriers, respectively. For example, among other results, detailed thermodynamic measurements of the Ba$_{1-x}$K$_{x}$Fe$_{2}$As$_{2}$ compound have shown a very low anisotropy of the second critical field, $\sim 3$~\cite{Ni2008,Wang2008c}.  The work performed on superconducting Ba(Fe$_{1-x}$Co$_{x})$As$_{2}$ has shown that Co doping does not introduce significant additional scattering~\cite{Sefat2008a}. The fishtail magnetization and weak anisotropy has also been studied.~\cite{Yamamoto2008} High-field properties~\cite{Ni2008a} as well as penetration depth studies~\cite{Gordon2008} of single crystals of this compound will be reported elsewhere. It is this system that we chose for the present study for its superior superconducting properties.

In this paper we use global and local magnetic properties as well as direct electro-transport measurements to study the details of vortex pinning and magnetic relaxation, evaluate supercurrent density, and finally construct the vortex phase diagram in single crystals of the recently discovered Fe-based pnictide superconductor, Ba(Fe$_{0.93}$Co$_{0.07}$)$_{2}$As$_{2}$.

We distinguish between the \textit{true} critical current density, $j_{c}$, which marks the crossover between flux flow and flux creep regimes, and the general supercurrent density, $j \leq j_c$, which is observed in the experiment. By definition of the critical current, at $j=j_{c}$ the current-dependent barrier for vortex escape from the pinning potential $U(j/j_c)=0$. The supercurrent is determined by the time-window of the experiment, $\Delta t$, and can be estimated from the logarithmic solution \cite{Geshkenbein1989} of the flux creep equation, $U\left(  j/j_{c}\right) = T\ln\left(1+\Delta t/t_{0}\right)$, where $t_{0} \sim 10^{-5}-10^{-7}$ s, is the \textit{macroscopic} characteristic time that depends on the sample size and shape as well as on the microscopic attempt time, $\tau_{0} \sim 10^{-10}-10^{-13}$ s.~\cite{Gurevich1993,blatter94,Burlachkov1998} (We use units where the energy barrier is measured in kelvin, so $k_{B}=1.$) It is actually the supercurrent that determines the usefulness of a particular superconductor for current-carrying applications and therefore the magnetic irreversibility should be analyzed both in terms of magnetic hysteresis and the rate of magnetic relaxation.

\section{Experimental}

Single crystals of Ba(Fe$_{0.93}$Co$_{0.07}$)$_{2}$As$_{2}$ were grown out of FeAs flux using high temperature solution growth techniques. More specifically, the powders of FeAs and CoAs were mixed with Ba in the ratio Ba:FeAs:CoAs=1:3.6:0.4. The mixture was placed into an alumina crucible and a second catch crucible containing quartz wool was placed on top of the growth crucible. Both were sealed in a quartz tube under argon and slowly heated to
1180 $^{\circ}$C, held for 2 hours, and slowly cooled to 1000 $^{\circ}$C over a period of 36 hours. Once the furnace has reached 1000 $^{\circ}$C the FeAs is decanted from the single crystals using a centrifuge. The size of the
resulting crystals can be as large as 12 x 8 x 1 mm$^{3}$. Elemental analysis was performed using wavelength dispersive X-ray spectroscopy in the electron probe microanalyzer of a JEOL JXA-8200 Superprobe. It showed the actual Co concentration is Co/(Co+Fe)=7.4\%, so the more accurate formula is Ba(Fe$_{0.926}$Co$_{0.074}$)$_{2}$As$_{2}$, but we round it to the second digit in the text. During the course of this study several samples from the same batch were measured. As shown below, they all have a transition temperature off $T_{c}=22$ K, as determined from dc magnetization (inset of Fig.~\ref{fig2}) and from transport measurements, Fig.~\ref{fig7}.

\begin{figure}[tb]
\begin{center}
\includegraphics[width=8.5cm]%
{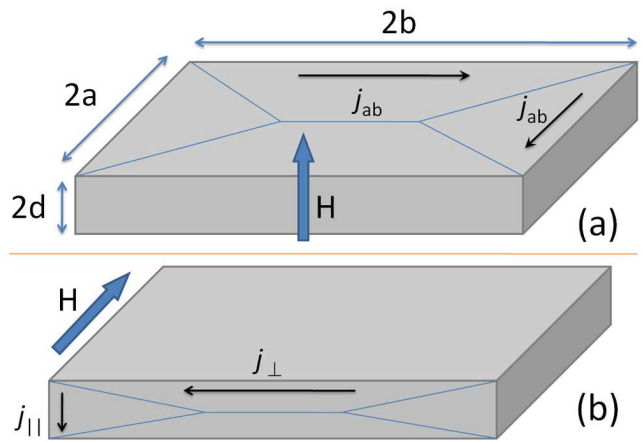}%
\caption{(Color online) Schematics used in calculations and definitions of the sample dimensions and flows of the supercurrents. (a) the magnetic field is applied perpendicular to the $ab-$ plane (parallel to the crystallographic $c-$ axis and $d-$ edge). (b) the magnetic field is applied along the $ab-$ plane (shown along the $a-$ edge).}%
\label{fig1}%
\end{center}
\end{figure}

For the measurements of total magnetic moment (global magnetic measurement), a cuboid-shaped sample (see Fig.~\ref{fig3}) of dimensions, $0.28 \times 0.7 \times 1.26$ mm$^3$, was fixed in a gelatine capsule with a small amount of Apiezon grease. The capsule was placed inside of a clear plastic straw. To check
for possible errors due to mechanical misalignment of the sample assembly, several key measurements (such as $M\left(  H\right)$ loops) were repeated three times each time after the sample was removed and re-assembled. No
noticeable variation in the results was found. The magnetization measurements were conducted in a \textit{Quantum Design} MPMS magnetometer. The second critical field, $H_{c2}\left(T\right)$, estimated from the onset of superconductivity, was measured by using a tunnel-diode resonator technique.~\cite{Martin2008}

\begin{figure}[tb]
\begin{center}
\includegraphics[width=8.5cm]%
{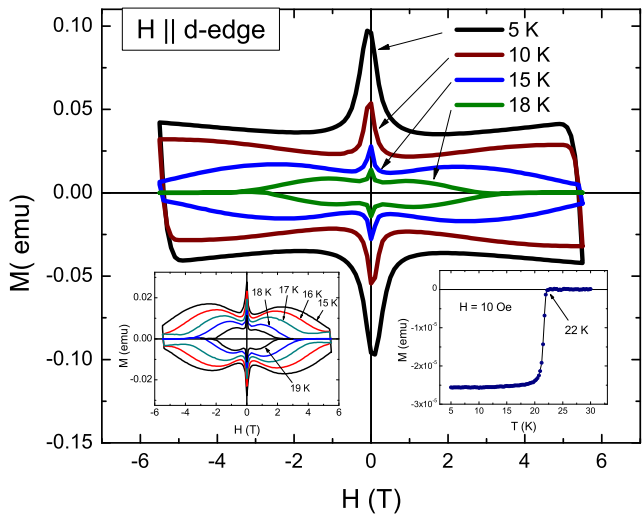}%
\caption{(Color online) Magnetization loops measured in a $0.28 \times 0.7 \times 1.26$ mm$^3$ single crystal of Ba(Fe$_{0.93}$Co$_{0.07}$)$_{2}$As$_{2}$ along the crystallographic $c-$ axis (along the $d-$ edge) at $T=5,~10,~15$ and $18$ K. Left inset zooms at the
higher temperature region showing the data for $T=15,~16,~17,~18$ and $19$ K where the evolution of fishtail magnetization is clearly seen. Right inset shows a superconducting transition measured after zero-field cooling (zfc) at $H=10$ Oe.}%
\label{fig2}%
\end{center}
\end{figure}

Samples for transport measurements were cut with a razor blade into long thin bars having typical dimensions of $(2-3) \times 0.1 \times 0.02$ mm$^3$. Contacts were made to the samples by soldering silver wires with a silver-based alloy and had negligible resistance (less than 0.1\% of the sample resistance).  This allowed for both two and four probe measurements to be taken. To sustain a high current density without thermal shock, samples were glued with GE-varnish onto an insulating heat sink substrate (\textit{LakeShore Cryotronics Inc.}). Current wires were thermally anchored to two silver foil heat sinks (see inset in Fig.~\ref{fig5}). Current-voltage characteristics, V(I), were measured in a \textit{Quantum Design} PPMS in constant current mode. The highest attainable current density was restricted to keep Joule heating less than 1 K at $T_c$. 

Magneto-optical (MO) imaging was performed in a $^{4}$He optical flow-type cryostat utilizing the Faraday rotation of polarized light in a Bi - doped iron-garnet indicator film with in-plane magnetization \cite{Prozorov2007d}.
The spatial resolution of the technique is about 3 $\mu$m with a sensitivity to magnetic fields of about 1 G. The temporal resolution, as determined by our image acquisition hardware, is about 30 msec. In all images, the intensity is proportional to the local value of the magnetic induction perpendicular to the sample surface.

\section{Results and discussion}

\subsection{Supercurrent density}

\subsubsection{Magnetization}

\begin{figure}[tb]
\begin{center}
\includegraphics[width=8.5cm]{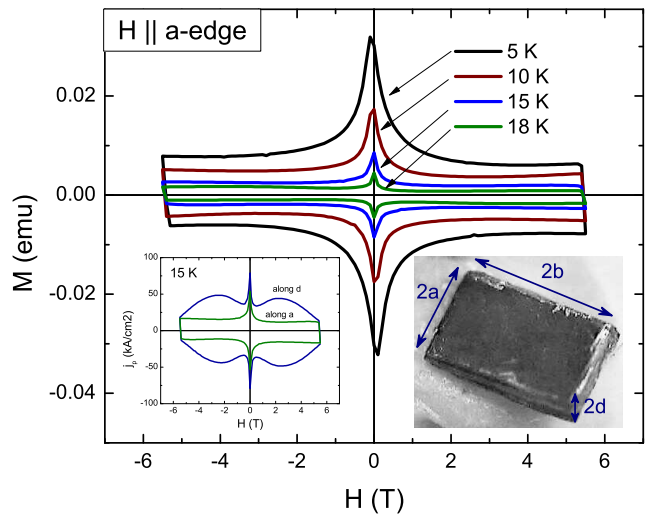}%
\caption{(Color online) Magnetization measured along the $a-$ edge in, perpendicular to the crystallographic $c-$ axis in a Ba(Fe$_{0.93}$Co$_{0.07}$)$_{2}$As$_{2}$ single crystal. Left inset: comparison between $M(H)$ loops at $T=15$ K in two orientations. Right inset: a photograph of the measured cuboidal sample with dimensions, $0.28 \times 0.7 \times 1.26$ mm$^3$.}%
\label{fig3}%
\end{center}
\end{figure}

Let us first define the geometry of the experiment. We consider a cuboid-shaped crystal with dimensions $2d<2a<2b$, as shown schematically in Fig.\ref{fig1} and one of the actual samples, $0.28 \times 0.7 \times 1.26$ mm$^3$, is shown in the inset of Fig.\ref{fig3}. In Ba(Fe$_{0.93}$Co$_{0.07}$)$_{2}$As$_{2}$, the crystallographic $ab-$ plane has the largest area and is parallel to the geometric $ab-$ plane of a cuboid. The smallest dimension $2d$ is the sample thickness. When a magnetic field is oriented along the crystallographic $c-$ axis (along the $d-$ edge), the measured magnetic moment per unit volume (volume magnetization) is denoted as $M_{d}$. Similarly, magnetic moments measured along the $b-$ edge and along the $a-$ edge are $M_{b}$ and $M_{a}$, respectively. Due
to tetragonal symmetry, the induced supercurrents can be considered isotropic in the $ab-$ plane (we use the designation $j$ for the in-plane supercurrent density), but may be different for the magnetic field oriented perpendicular to the $c-$axis. For the vortex motion crossing the Fe-As planes, the shielding current density is $j_{\perp}$ and for vortices moving parallel to these planes we use $j_{\parallel}$. Details of the anisotropic response will be published elsewhere.~\cite{Tanatar2008}%

Figure \ref{fig2} shows $M\left(  H\right)  $ loops measured at several temperatures. A noticeable nonmonotonic "fishtail" magnetization develops at elevated temperatures. (It presumably exists also at the lower temperatures, but is shifted to higher magnetic fields beyond the capabilities of our setup). The left inset in Fig.\ref{fig2} zooms into this temperature interval to show the details of the fishtail evolution with temperature. This behavior is
quite similar to YBCO single crystals.~\cite{Panetta1995,Abulafia1996}

The situation is quite different when magnetization is measured with a magnetic field applied along the crystallographic $ab-$ plane. Figure~\ref{fig3} shows measurements performed with $H$ along the $a-$ edge of the
crystal. No noticeable fishtail behavior is observed. The inset compares $M(H)$ loops at $T=15$ K for the two orientations where the difference is clearly seen. When a magnetic field was applied along the longer  $b-$ edge,
the result  was the same as in the measurement along the $a-$ edge but with a different magnitude due to the difference in the cross-section perpendicular to the field. Similar behavior of $M\left(  H\right)  $ in two
orientations has been observed in several crystals with different aspect ratios varying by orders of magnitude, which excludes possible geometric effects.

\begin{figure}[tb]
\begin{center}
\includegraphics[width=8.5cm]{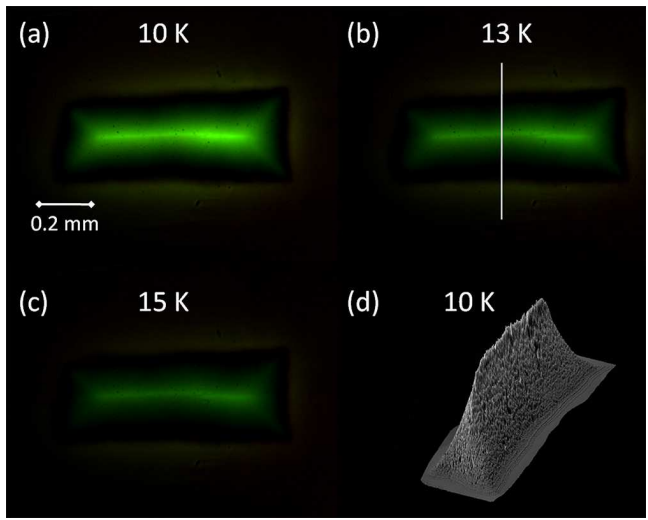}%
\caption{(Color online) Magneto-optical images of a single crystal Ba(Fe$_{0.93}$Co$_{0.07}$)$_{2}$As$_{2}$ ($0.053 \times 0.24 \times 0.55$ mm$_3$) in the remanent state. (Magnetic field was applied along the $c-$axis (perpendicular to the page) above $T_c$ and removed after cooling to 5 K). Frames (a), (b) and (c) show images for $T=$ 10, 13 and 15 K, respectively. Frame (d) shows a three-dimensional Bean oblique wedge where the $z-$ axis is intensity corresponding to the frame (a).}%
\label{fig4}%
\end{center}
\end{figure}

\subsubsection{Magneto-optical imaging}

In order to evaluate the supercurrent density we need to examine the structure of the critical state by observing the distribution of the magnetic induction in the sample. The magneto-optical images of a sample from the
same batch used in $M(H)$ measurements are shown in Fig.\ref{fig4}. The sample dimensions were $0.053 \times 0.24 \times 0.55$ mm$_3$. Frames (a), (b), and (c) show the remanent state obtained at $T=$ 10, 13, and 15 K, respectively. Figure \ref{fig4} (d) shows a 3D plot where the $z-$ axis is the magnetic induction.  This oblique wedge shape is what is expected from the Bean critical state model~\cite{Bean1962,Bean1964}. We can therefore use this well-known approach to calculate the supercurrent density from the measured magnetization and the profiles of magnetic induction across the sample.%

\begin{figure}[tb]
\begin{center}
\includegraphics[width=8.5cm]{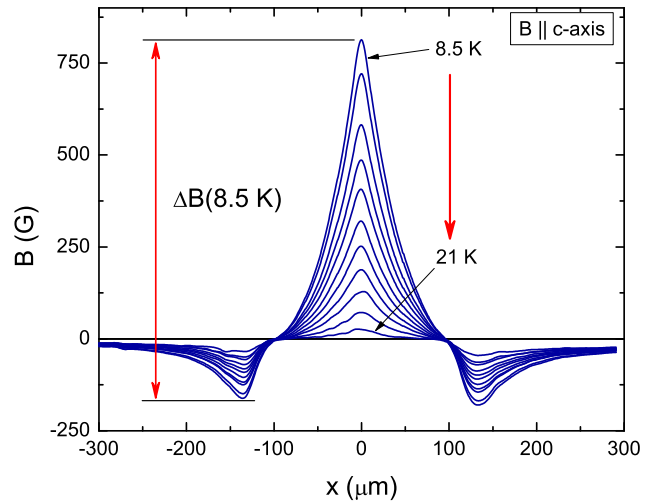}%
\caption{(Color online) Profiles of the magnetic induction measured in a single crystal Ba(Fe$_{0.93}$Co$_{0.07}$)$_{2}$As$_{2}$ at different temperatures along the line shown in Fig.~\ref{fig4}(b). Definition of the maximum variation of the magnetic induction, $\Delta B$ is shown for profile obtained at $T=8.5$ K.}%
\label{fig5}%
\end{center}
\end{figure}

We can also examine the profiles of magnetic induction, $B(r)$, across the sample obtained along the line shown in Fig.~\ref{fig4}(b). Figure \ref{fig5} shows such profiles obtained from the measurements at different temperatures. The shape of the profile is typical for a thin slab-like sample. There is a clear neutral line due to self-fields that are generated by the Bean supercurrents. A graphical definition of the maximum variation of $B_{z}\left(  r\right)  $, used later for the evaluation of the supercurrent density, is shown for the $T=8.5$ K profile, as an example.

\subsubsection{Evaluation of the supercurrent density from various measurements}

\begin{figure}[b]
\begin{center}
\includegraphics[width=8.5cm]{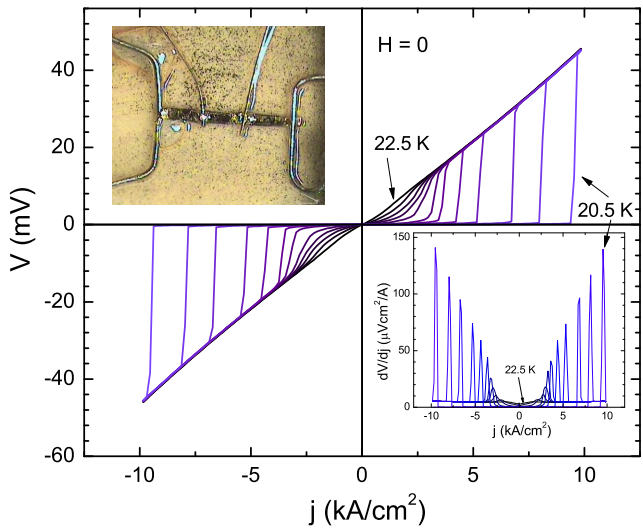}%
\caption{(Color online) Voltage as a function of current density measured at different temperatures in a bar-shaped, $3 \times 0.1 \times 0.02$ mm$^3$, single crystal of Ba(Fe$_{0.93}$Co$_{0.07}$)$_{2}$As$_{2}$ with current applied along the $ab-$ plane. Upper inset: A photograph of the sample with contacts. Lower inset: The derivative $dV/dj$ used to identify the critical current, $j$.}%
\label{fig6}%
\end{center}
\end{figure}

We now extract the supercurrent density from global and local magnetic measurements and compare it to direct transport measurements  that were performed on the crystals selected from the same batch. Current-voltage characteristics are shown in Fig.~\ref{fig6}. In a single crystal that was still quite thick for transport measurements, we could only induce the transition to the normal state in a very limited temperature interval. This was sufficient to overlap with the magnetization data to check whether our indirect calculations of $j$ result in a comparable magnitude of the supercurrent. Sample dimensions were determined in a calibrated optical microscope and we estimate the (systematic) errors on the order of 10\% for the evaluated supercurrent density. Random errors were negligible.

\begin{figure}[tb]
\begin{center}
\includegraphics[width=8.5cm]{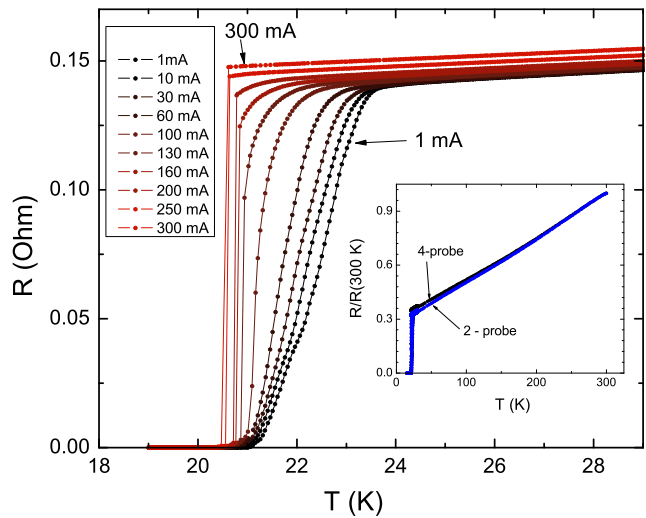}%
\caption{(Color online) Resistance as a function of temperature for different values of a dc transport current in a $3 \times 0.1 \times 0.02$ mm$^3$ single crystal of Ba(Fe$_{0.93}$Co$_{0.07}$)$_{2}$As$_{2}$. The onset of non-zero resistance was used to determine the supercurrent density. Inset: normalized resistance in a full-temperature range measured by two- and four- probe methods.}%
\label{fig7}%
\end{center}
\end{figure}

Figure \ref{fig6} shows current - voltage, $V(j)$, characteristics measured at different temperatures. The Inset of Fig.~\ref{fig6} shows a photograph of the sample with the contacts. The temperature dependence of the resistance measured at different excitation currents is shown in Fig.~\ref{fig7}.  The resistance, normalized by the room-temperature value, measured by a two-probe technique between current leads and in the conventional four-probe configuration in the full temperature range, is shown in the inset of Fig.~\ref{fig7}. The data scales perfectly with the sample geometry, as the distance between contacts for the voltage readings are different.

\begin{figure}[tb]
\begin{center}
\includegraphics[width=8.5cm]{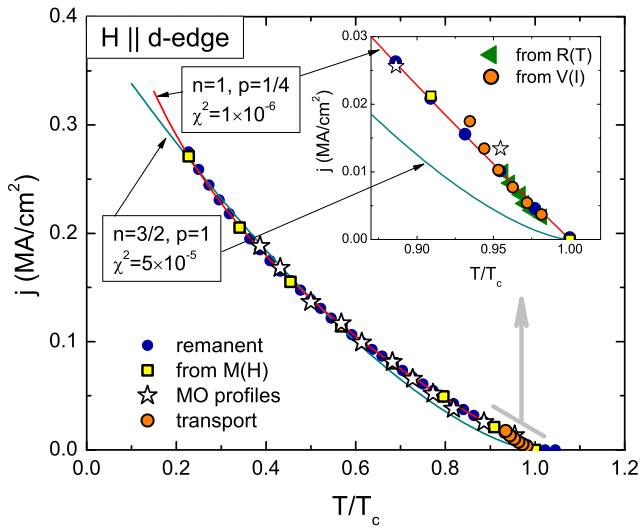}%
\caption{(Color online) Supercurrent density in same-batch single crystals Ba(Fe$_{0.93}$Co$_{0.07}$)$_{2}$As$_{2}$. "Magnetic" current density estimated from Eq.~\ref{jBean} at zero applied field by annealing of the remanent magnetization (filled circles), from $M(H)$ loops measured at different temperatures (squares), from magneto-optical profiles using Eq.~\ref{DB} (stars), and from direct transport measurements (open circles). The lines show fits to $j(T)=j(0)\left( 1- \left( T/T_c \right)^p \right)^n$ of the dataset shown in full circles with indicated exponents and adjusted $\chi^2$ values. Inset zooms at the region of $T_c$ for easier comparison with the transport data obtained from Fig.~\ref{fig6} (circles) and Fig.~\ref{fig7} (triangles).}%
\label{fig8}%
\end{center}
\end{figure}

In the normal state the $V(j)$ curves remain linear up to $j \approx 10$ kA/cm$^2$. For higher $j$, self heating effects raise the slope of the curve. Below $T_c$, a notable curvature in $V(j)$ develops at zero bias and eventually a broad region of zero voltage appears at the lower temperatures indicating a true superconducting state. With the increase of the current density, the voltage crosses over to a linear dependence as it should be in the normal state. From these measurements we have determined the value of the critical current density as the point at which the derivative, $dV/dj$, is maximal as shown in Fig.~\ref{fig6}. Though this definition does not represent the true zero resistivity state in the vicinity of $T_c$, from below the transition this corresponds closely to a point of sharp increase in the measured voltage and is easy to determine. In order to avoid dynamic effect related to sweeping the current, we also estimated the current density from $R(T)$ curves measured at different values of the bias current. The $R(T)$ measurements are shown in Fig.~\ref{fig7} and the extracted current densities are compared to those obtained from $V(I)$ and magnetic measurements in the inset Fig.~\ref{fig8}. The overall temperature dependence of the supercurrent density obtained at different temperatures by different techniques are summarized in Fig.~\ref{fig8}.

To determine the supercurrent from magnetic measurements, we use the Bean model in which there is a field-independent $j$.~\cite{Bean1962,Bean1964} Assuming that $j$ is isotropic in the crystallographic $ab-$ plane, there are three different current densities depending on the orientation of an external magnetic field with respect to the crystal faces. When the magnetic field is oriented along the $d-$ edge (crystallographic $c-$axis), Abrikosov vortices move in the $ab-$ plane and gradients in their density induce the supercurrent, $j$, given by 

\begin{equation}
j_{ab}=\frac{c M_{d}}{a}\left(  1-\frac{a}{3b}\right)^{-1}
\label{jBean}
\end{equation}

\noindent The other two measured component of volume magnetization, $M_{a}$ and $M_{b}$, can be used to calculate $j_{\parallel}$ and $j_{\perp}$. A detailed study of the anisotropic properties, including supercurrents, will be published elsewhere.~\cite{Tanatar2008}

Alternatively, the supercurrent can be calculated directly from the magnetic induction profiles, $B_{z}\left(  r\right)  $, measured along the line shown in Fig.~\ref{fig4} (b). It is worth noting that the total magnetization and
magneto-optics are two very different measurements performed under different experimental conditions. The only common part of the analysis is the Bean model used for data analysis. If this assumption is invalid, the resulting supercurrent densities will be very different. The variation of magnetic induction, $\Delta B=\left\vert B_{z}\left(0\right)  -B_{z}\left(  \text{edge}\right)  \right\vert $ in the full critical state is given by \cite{PhDProzorov1998}

\begin{equation}
\frac{c \Delta B}{4j a}={\eta\ln\frac{{\left(
{1+4\eta^{2}}\right)  ^{2}}}{16\eta^{3}\sqrt{1+\eta^{2}}}+2\arctan\left(
{2\eta}\right)  -\arctan\left(  \eta\right)  }
\label{DB}
\end{equation}

\noindent where $\eta=d/a$. In the sample used for MO study, $\eta \approx 0.22$ and Eq.~\ref{DB} yields $j \approx 194 \Delta B$. Note that if instead of using Eq.~\ref{DB} one would use the straightforward Bean model for a semi-infinite slab (we still measure the $z-$ component of $\mathbf{B}$ on the surface), then the estimated current density would be significantly lower, $j=c/(2 \pi a) \Delta B \approx 133 \Delta B$.

We first analyze the supercurrent density for zero applied field. Figure \ref{fig8} shows the supercurrent density estimated from magnetization, magneto-optical, and direct transport measurements. From magnetization,
the current was obtained from two types of measurements. First, a full remanent state was induced at 5 K by cooling in a 5 T magnetic field and then turning the field off. Then, this remanent state was slowly warmed up, allowing the temperature to stabilize before the measurement was taken. This convenient method is relatively quick and produces a curve with many data points. However, it is unclear whether we probe the original critical state relaxed only during the time window $\Delta t$, characteristic of our magnetometer, or the supercurrent has relaxed more and we are probing a deeply relaxed state. Therefore, we have used $M(H=0)$ values directly from the magnetization loops measured at different
temperatures and plot the resulting $j$ as open squares in Fig.\ref{fig8}. The data simply falls on top of the curve obtained from the annealing of the remanent state (filled circles). The explanation is that the change of temperature occurs during a relatively short period of time, comparable to the time of setting and stabilizing the magnetic field in the superconducting magnet of the MPMS magnetometer during the field ramp. The supercurrent density evaluated from Eq.\ref{DB} (shown by stars in Fig.~\ref{fig8}) compares well with the magnetization data. Finally, the inset in Fig.~\ref{fig8} compares the supercurrent obtained from magnetic measurements with direct transport measurements. There is good agreement between all datasets. We therefore conclude that, similar to NdFeAsO$_{1-x}$F$_{x}$~\cite{Fe_Prozorov2008}, the supercurrent density in single crystals Ba(Fe$_{0.93}$Co$_{0.07}$)$_{2}$As$_{2}$ is relatively low even at low temperatures, $j\left(  T=5\text{K},H=0\right) \simeq 2.6 \pm 0.2 \times 10^{5}$ A/cm$^{2}$, which is comparable to Y-Ba-Cu-O single crystals. As for the functional form of $j_c(T)$, we can use the generalized power law that is usually used for superconductors, $j(T)=j(0)\left( 1- \left( T/T_c \right)^p \right)^n$, which gives the best fit to the experimental data (most data points - solid symbols) with $j(0) \approx 0.88$ MA/cm$^2$, $p=1/4$ and $p=1$. An attempt to fit with $p=1$ and $n=3/2$ is also shown in Fig.~\ref{fig8} and is inferior to the former fit, which is reflected in an order of magnitude larger adjusted $\chi^2$ value.
\begin{figure}[tb]
\begin{center}
\includegraphics[width=8.5cm]{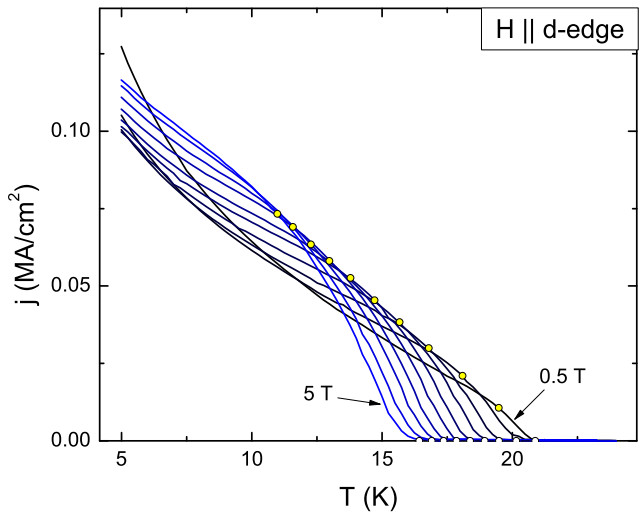}%
\caption{(Color online) $j\left( T \right)$ measured in a single crystal Ba(Fe$_{0.93}$Co$_{0.07}$)$_{2}$As$_{2}$ upon warming up at different values of the magnetic applied field as explained in the text. Open circles mark the temperature of the vanishing current as well as the temperatures at which $j\left(T\right)$ curves exhibit an apparent change of behavior.}%
\label{fig9}%
\end{center}
\end{figure}
Next we present unexpected results obtained in a magnetic field. Using the annealing method described above for the remanent state, but this time annealing the induced critical state in finite fields, we have obtained a set of
$j\left(  H=const,T\right)$ curves for different values of $H$. The result is shown in Fig.~\ref{fig9}. There is an apparent pronounced decrease in the current density marked by open circles. Similarly, the temperature at which
$j_{ab}\rightarrow 0$ is also marked. While the meaning of the former signature will become clear below when we discuss the $H-T$ diagram, the latter feature is obviously the experimental irreversibility line above which
pinning is negligible and the vortex liquid sets in.

\subsection{Magnetic relaxation}

When the fishtail effect is observed, one must examine the magnetic relaxation of the critical state as a function of temperature and magnetic field. While some theories suggest a static explanation for the fishtail effect, which most likely works in alloys, ceramics, and materials with secondary phases, in single crystals it seems that a dynamic scenario is more plausible, at least in the high-$T_{c}$ cuprates.~\cite{Yeshurun1996,Burlachkov1998} Within collective pinning and creep models, the magnetic relaxation rate \textit{decreases} with the increase of a magnetic field. Since the true critical current density, $j_{c}\left( H\right)$, also \textit{decreases} with the increase of a magnetic field, an apparent fishtail develops in the measured magnetization.~\cite{Burlachkov1998} Therefore, the fishtail phenomenon is a direct consequence of the collective pining and creep theory. However, more detailed measurements above the peak position have forced us to revisit this picture and use a refined model that involves both field-dependent relaxation rates and a transition from a collective flux creep mechanism at the lower fields to a plastic creep (mediated by dislocations in the vortex lattice).~\cite{Abulafia1996} The latter predicts an \textit{increasing} relaxation rate with the increase of the magnetic field, whereas the former predicts a monotonic decrease of the relaxation rate.

\begin{figure}[tb]
\begin{center}
\includegraphics[width=8.5cm]{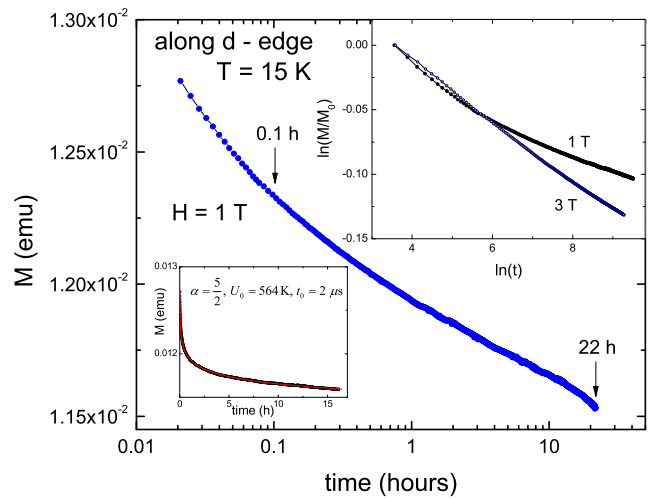}%
\caption{(Color online)Time - dependent magnetization measured for 22 hours at $T=15$ K at $H=1$ T after ramping field from $5$ T. Lower inset: fit to the collective creep model, Eq.~\ref{M(t)}, with parameters indicated. Upper inset: $\ln-\ln$ scale graph of relaxation curves obtained at $1$ T and $3$ T.}%
\label{fig10}%
\end{center}
\end{figure}

Based on the various models of magnetic relaxation, Griessen \textit{et al.} has suggested a very useful generic form of the barrier for flux creep.~\cite{Griessen1997}
\begin{equation}
U\left(  j\right)  =\frac{U_{0}}{\alpha}\left[  \left(  \frac{j_{c}}%
{j}\right)  ^{\alpha}-1\right]
\label{U(j)}
\end{equation}

This formula describes all the widely-accepted functional forms of $U\left(j/j_c\right)$ if the exponent $\alpha$ is allowed to have both negative and positive values. For $\alpha=-1$, Eq. \ref{U(j)} describes the classical
Anderson-Kim barrier, which is simply linear in $j/j_{c}$ \cite{Anderson1962,Anderson1964}. For $\alpha=-1/2$
the barrier for plastic creep~\cite{Abulafia1996} is obtained. Positive $\alpha$ describes collective creep barriers \cite{blatter94}. In the limit $\alpha \rightarrow 0$, this formula reproduces exactly the logarithmic barrier.~\cite{Zeldov1989} An activation energy, written in the form of Eq.~\ref{U(j)}, results in a so-called interpolation formula for flux creep~\cite{blatter94} obtained when the logarithmic solution of the creep equation $U\left(j/j_c\right) = T\ln(1+t/t_{0})$~\cite{Geshkenbein1989} is applied (for $\alpha \neq 0$),
\begin{equation}
M\left(  t\right)  =M_{c}\left(  1+\frac{\alpha T}{U_{0}}\ln\left(  \frac
{t}{t_{0}}\right)  \right)  ^{-\frac{1}{\alpha}}
\label{M(t)}%
\end{equation}

For $\alpha=0$, a power-law decay is obtained: $M\left(  t\right)=M_{c}\left(  t_{0}/t\right)  ^{n}$, where $n=T/U_{0}$. A generalization of the flux creep theory beyond the logarithmic solution with applications to
collective creep and the fishtail effect are given by Burlachkov, Giller, and Prozorov.~\cite{Burlachkov1998}

Indeed, no single equation for relaxation is applicable for all values of $j/j_{c}$. Within the collective creep theory, the magnetic relaxation has different functional forms (roughly described by using different $U_{0}$ and $\alpha$ in Eq.\ref{U(j)}) depending on whether one considers the regime of a single-vortex, small vortex bundles, or large ones~\cite{blatter94}. Trying to find the best fit to Eq.~\ref{M(t)} in the widest time-interval, we have determined that the small-bundle regime describes the data shown in Fig.~\ref{fig10} quite well. The fit is shown in the lower inset. The fit parameters are reasonable within the collective creep model, as $\alpha=5/2$ (fixed during fitting), $U_{0}=564$ K, and $t_{0}=2$ $\mu s$.

\begin{figure}[tb]
\begin{center}
\includegraphics[width=8.5cm]{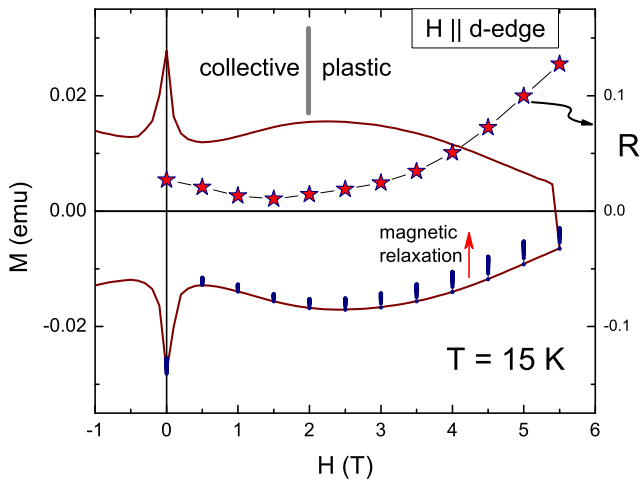}%
\caption{(Color online) Magnetic relaxation measured in a single crystal Ba(Fe$_{0.93}$Co$_{0.07}$)$_{2}$As$_{2}$  at different values of the applied magnetic field shown along with the $M(H)$ loop measured at the same temperature, $T=15$ K. Right axis shows the logarithmic relaxation rate as function of an applied field.}%
\label{fig11}%
\end{center}
\end{figure}

We now examine relaxation magnetic fields above and below the peak position, $H_p(T)$. As can be seen in the upper inset of Fig.~\ref{fig10}, already just above the peak, the relaxation rate is faster than before the peak. Its functional form is also quite different. The magnetic field dependence on the relaxation rate is presented in Fig.~\ref{fig11}. Here, each $M\left(  H=const,t\right)$ trace is plotted along with the regular magnetization loop.%

A useful characteristic of the flux creep is the logarithmic relaxation rate, $R=-\left\vert d\ln M/d\ln t\right\vert $. From Eq.\ref{M(t)} we obtain, 

\begin{equation}
R=\frac{T}{U_{0}}\left(  1+\frac{\alpha T}{U_{0}}\ln\left(  \frac{t}{t_{0}%
}\right)  \right)  ^{-1}%
\label{R}%
\end{equation}

Even at the longest time that we have probed (22 hours), the logarithmic term is equal to $1.6$, so the relaxation rate only decreases by a factor of about $2.6$ compared to the initial (but not too close to $t=0$) rate, $R\simeq T/U_{0}$,
which at $15$ K and $1$ T is about $R=0.027$. This value is much larger than the rate for conventional superconductors, but is quite similar to high-$T_{c}$ compounds.~\cite{Yeshurun1996} Thus, single crystals of Ba(Fe$_{0.93}$Co$_{0.07}$)$_{2}$As$_{2}$ exhibit "giant flux creep". The magnetic field dependence of $R$ is summarized in Fig.~\ref{fig11} (right axis). At first, it decreases with
the increase of $H$, up to the field, $H_{p}$, corresponding to the maximum in the fishtail $M\left(H\right)$ curve. Above $H_{p}$ the relaxation rate increases with the increase of $H$, which contradicts the collective creep
theory. To clarify whether this behavior is related to the fishtail phenomenon, we have measured the magnetic relaxation rate with the magnetic field applied along the $a-$ edge and we did not observe a fishtail feature (Fig.~\ref{fig3}).

\begin{figure}[tb]
\begin{center}
\includegraphics[width=8.5cm]{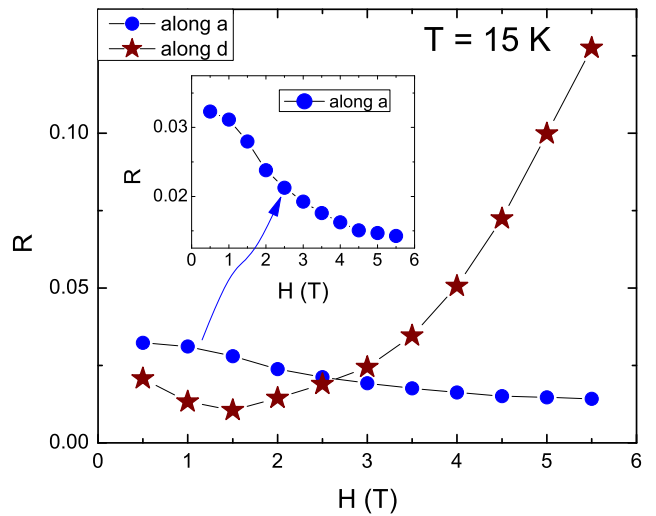}%
\caption{(Color online) Comparison of the relaxation rate, $R$, as function of an applied magnetic field measured in a single crystal Ba(Fe$_{0.93}$Co$_{0.07}$)$_{2}$As$_{2}$ in two different orientations at $T=15$ K.}%
\label{fig12}%
\end{center}
\end{figure}

Figure \ref{fig12} compares the relaxation rates in the two orientations. While above $H_{p}$, $R\left(  H\right)  $ increases with $H$ for measurements along the $c-$ axis, it shows the opposite trend when a magnetic field is applied along the $ab-$ plane. We therefore conclude that the collective creep model fails above $H_{p}$.%

\begin{figure}[b]
\begin{center}
\includegraphics[width=8.5cm]{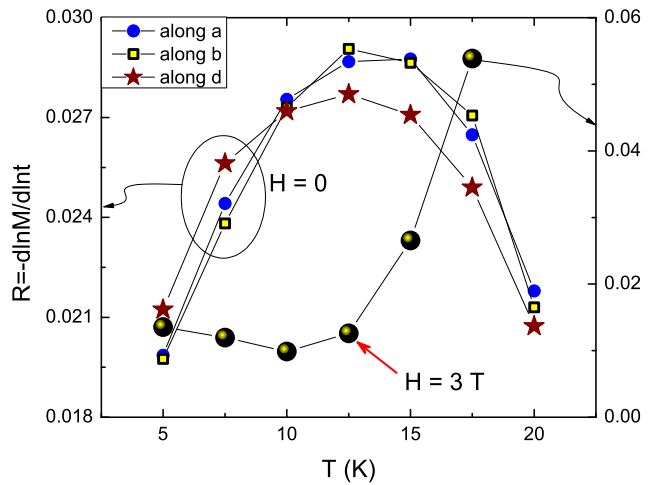}%
\caption{(Color online) Logarithmic relaxation rate as function of temperature measured in a single crystal Ba(Fe$_{0.93}$Co$_{0.07}$)$_{2}$As$_{2}$ in the remanent state, $H=0$, in three orientations of the trapped vortex direction with respect to the sample. For comparison, $R\left(  T\right)  $ at $H=3$ T is shown. (Note different origins and scales of the left and right
axes).}%
\label{fig13}%
\end{center}
\end{figure}

Another difference between the relaxation rates comes from examining the temperature dependence of $R$. Figure \ref{fig13} compares $R\left(  T\right)$ measured in zero field (self-field of the remanent state) for three different
orientations (along three of the edges of the crystal). When vortices are parallel to the $ab-$ plane, the relaxation rates are practically the same. For magnetic relaxation in the remanent state obtained along the $c-$ axis, $R\left(  T\right)$ shows a similar functional dependence and is shifted slightly to lower temperatures. In contrast, the relaxation rate obtained above $H_{p}$ shows a very different behavior - increasing with the increase of temperature. The bell-shaped $R\left(  T\right)  $ are observed in many superconductors with weak pinning. No straightforward explanation exists, but it seems plausible that while the bare barrier for the magnetic relaxation, $U_{T,j=0}$, decreases with the increase of temperature, the ratio $j/j_{c}$ at which the relaxation takes place in the experiment with the fixed time window $\Delta t$ decreases with the increase of temperature. Therefore, the actual barrier experienced by the vortex bundles, Eq.~\ref{R}, may become non-monotonic with temperature. In other words, the exponent $\alpha$ is temperature dependent and from the collective creep theory it follows that it is nonmonotonic with the bundle size, peaking at $\alpha=5/2$ for small bundles \cite{blatter94}. By substituting this into Eq.~\ref{R} with the appropriate $U_{T,j=0}\left(  T\right)$, the nonmonotonic $R\left(  T\right)  $ can be reproduced. This provides additional indirect evidence for the applicability of the collective creep approach to vortex relaxation in single crystalline Ba(Fe$_{0.93}$Co$_{0.07}$)$_{2}$As$_{2}$ at fields below $H_{p}$. Using the same arguments, one can show that plastic creep, which corresponds to $\alpha=-1/2$ cannot lead to such behavior and the relaxation rate should just increase with temperature, which is what we observe in Fig.~\ref{fig13}.

The idea of plastic vortex creep was introduced when a similar failure of the collective creep model was found in YBCO\cite{Abulafia1996} and the electron-doped cuprate superconductor, Nd$_{0.85}$Ce$_{0.15}$CuO$_{4-x}$ (NCCO)~\cite{Giller1997}, single crystals. Similar measurements were used to extract $E\left( j\right)$ characteristics without attaching contacts and confirmed the plastic creep model~\cite{Giller1998}. These and later works
showed that in superconducting crystals with relatively weak pinning, the crossover from elastic to plastic creep always accompanies fishtail magnetization. However, it should be noted that the fishtail feature itself is still due to peculiarities of
the collective creep when the relaxation rate is faster for lower fields, $H < H_p$. Within the plastic creep scenario, thermally activated vortex motion is not due to jumps of vortex bundles, but due to sliding of dislocations in the vortex lattice (even in a very disordered state, one can consider a local ordered arrangement of vortices and dislocations as primary defects). The main result of this approach is that the barrier for magnetic relaxation in the plastic creep regime remains finite as $j \rightarrow 0$, $U_{pl}\left(  B,j/j_{c}\right)
=U_{pl}\left(  B\right)  \left(  1-\sqrt{j/j_{c}}\right)  $ and decreases with increasing magnetic field as $U_{pl}\left(  B\right)  \sim 1/\sqrt{B}$, because the elementary movement of a dislocation is proportional to the
intervortex distance, $a \sim 1/\sqrt{B}$. Collective creep barriers, on the other hand, increase with increasing magnetic field as $U_{c}\left(  B\right)  \sim B^{\nu}$ , where $\nu$ is positive and depends on the particular pinning regime. The $j$ dependence is given by Eq.\ref{U(j)} with a positive $\alpha$ that results in a diverging barrier, $U_{c}\left(  j/j_{c}\right)  \sim\left(j/j_{c}\right)  ^{\alpha}$. At low fields, the collective barrier is always smaller than the plastic barrier and this channel dominates the vortex dynamics. However, as the magnetic field is increased, a crossover from a collective to a plastic channel of vortex relaxation occurs when
$U_{pl}\left(  B,j/j_{c}\right)  $ becomes smaller than $U_{col}\left(  B,j/j_{c}\right)$.

\begin{figure}[tb]
\begin{center}
\includegraphics[width=8.5cm]{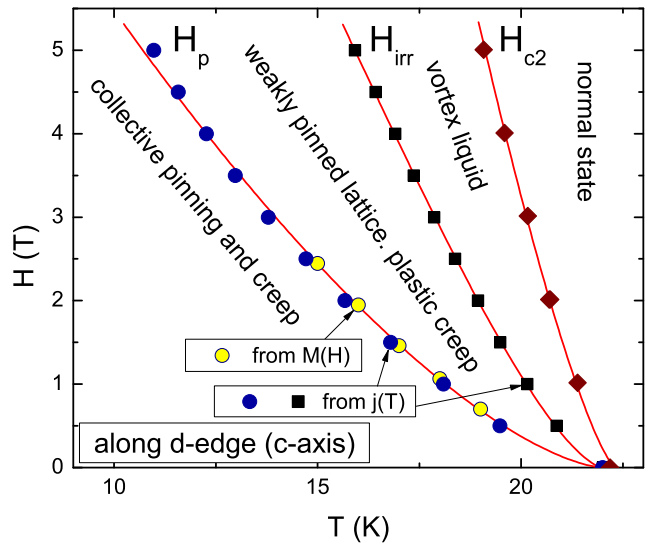}%
\caption{(Color online) Vortex $H-T$ phase diagram obtained from various features in the described measurements and the solid lines are the fits as discussed in the text.}%
\label{fig14}%
\end{center}
\end{figure}

Finally, Fig.~\ref{fig14} presents the vortex phase diagram of Ba(Fe$_{0.93}$Co$_{0.07}$)$_{2}$As$_{2}$ single crystals compiled from our results. It closely resembles diagrams for the cuprate superconductors, in particular YBCO single crystals. The prominent features are the irreversibility line that is quite distant from the $H_{c2}\left(  T\right)$ line and a clear crossover line between collective and plastic creep regimes. As is usually the case, the data can be described by the general power law, $H_{xx}(T)=H_{xx}(0)\left( 1- \left( T/T_c \right)^p \right)^n$, shown in Fig.~\ref{fig14} by solid lines. We obtained for the fishtail peak position, $H_p(T)$, $H_p(0) \approx 13.6$ T, $n=3/2$, and $p=1$. For the irreversibility line, $H_{irr}(T)$, $H_{irr}(0) \approx 15.2$ T, $n=3/2$, and $p=2$. For the second critical field, $H_{c2}(T)$, $H_{c2}(0) \approx 66.7$ T, $n=4/3$, and $p=1$.

\section{Conclusions}

In conclusion, detailed measurements of global and local electromagnetic properties clearly show that the vortex behavior in single crystalline superconducting Ba(Fe$_{0.93}$Co$_{0.07}$)$_{2}$As$_{2}$ is similar to that found in the high $T_{c}$ cuprates. In particular, we report the presence of a fishtail feature in $M\left(  H,T=const\right)$ loops and find its signature in $M\left(  H=const,T\right)$ measurements. Furthermore, magnetic relaxation measurements are consistent with the collective pinning and creep models (weak pinning and fast creep) and suggest a crossover into the plastic creep regime in fields exceeding the value corresponding to the maximum in fishtail magnetization.

\begin{acknowledgments}
Acknowledgements: We thank J. Clem, A. Koshelev, R. Mints and J. Schmalian for helpful discussions. Work at the Ames Laboratory was supported by the Department of Energy-Basic Energy Sciences under Contract No.DE-AC02-07CH11358. R. P. acknowledges support from Alfred P. Sloan Foundation.
\end{acknowledgments}

\end{document}